\documentclass[12pt]{article}
\usepackage{graphicx}
\usepackage{epstopdf}
\usepackage{epsfig}
\usepackage{amsmath}
\usepackage{multirow}
\usepackage{subfigure}
\usepackage{rotating}
\usepackage{array}
\def\lsim{\mathrel{\raise.3ex\hbox{$<$\kern-.75em\lower 1ex\hbox{$\sim$}}}}
\def\gsim{\mathrel{\raise.3ex\hbox{$>$\kern-.75em\lower 1ex\hbox{$\sim$}}}}

\def\be{\begin{equation}}
\def\ee{\end{equation}}
\def\bea{\begin{eqnarray*}}
\def\eea{\end{eqnarray*}}

\begin{document}
\title{Dual models of the neutrino mass spectrum}
\author{Jiajun Liao$^1$, D. Marfatia$^{2,3,4}$, and K. Whisnant$^1$\\
\\
\small\it $^1$Department of Physics and Astronomy, Iowa State University, Ames, IA 50011, USA\\
\small\it $^2$Department of Physics and Astronomy, University of Hawaii at Manoa, Honolulu, HI 96822, USA\\
\small\it $^3$Department of Physics and Astronomy, University of Kansas, Lawrence, KS 66045, USA\\
\small\it $^4$Kavli Institute for Theoretical Physics, University of California, Santa Barbara, CA 93106, USA}
\date{}
\maketitle

\begin{abstract}

We show that any model with a homogeneous relationship among elements
of the neutrino mass matrix with one mass hierarchy yields predictions
for the oscillation parameters and Majorana phases similar to those
given by a model with the same homogeneous relationship among
cofactors of the neutrino mass matrix with the opposite mass
hierarchy, except when the lightest mass is of order 20 meV or less.

\end{abstract}

\newpage

The study of neutrino physics has greatly progressed in the last
twenty years. On the one hand, there have been many experimental
discoveries and breakthroughs that have culminated in determinations
of the two mass-squared differences between the three neutrinos and
the three neutrino mixing angles.  On the theoretical side, neutrino
models have been built to explain the experimental results and provide
guidance for the next generation of experiments.

Many models predict a relationship among the elements of the light
neutrino mass matrix, while other models predict the same relationship
among the cofactors of the light neutrino mass matrix. In
Refs.~\cite{Liao:2013kix, Liao:2013rca}, we found that there are
strong similarities between single texture zero models with one mass
hierarchy and single cofactor zero models with the opposite mass
hierarchy if the lightest mass in each case is not too small. This
curious feature was also discussed in Ref.~\cite{Lashin:2011dn}. The
phenomenon is not unique -- models with two equalities between mass
matrix elements are similar to models with two equalities between
cofactors with the opposite mass hierarchy, as noted in
Ref.~\cite{Dev:2013xca}. We find~\cite{inprep} this similarity to also
exist between models with two texture zeros in the light neutrino mass
matrix~\cite{Frampton:2002yf} and models with two cofactor zeros in
the light neutrino mass matrix~\cite{Lavoura:2004tu}.

In this article we generalize this correspondence by showing that any
model with a homogeneous relationship among elements of the light
neutrino mass matrix with one mass hierarchy predicts oscillation
parameters and Majorana phases similar to those of models with the
same homogeneous relationship among cofactors of the mass matrix with
the opposite mass hierarchy. Since the neutrino mass hierarchy remains
undetermined, two such models are indistinguishable using current
data. The allowed oscillation parameters are nearly identical when the
masses are quasi-degenerate, but can differ in some cases when the
lightest neutrino mass is of order 20 meV or less.

{\bf Comparison between element and cofactor models.}
The neutrino mass matrix can be written as 
\begin{equation}
M=V^*\text{diag}(m_1, m_2, m_3)V^\dagger\,,
\label{eq:Mnu}
\end{equation}
where $V=U\text{diag}(1, e^{i\phi_2/2}, e^{i\phi_3/2})$ is unitary
and~\cite{PDG}
\begin{align}
U=\begin{bmatrix}
   c_{13}c_{12} & c_{13}s_{12} & s_{13}e^{-i\delta} \\
   -s_{12}c_{23}-c_{12}s_{23}s_{13}e^{i\delta} & c_{12}c_{23}-s_{12}s_{23}s_{13}e^{i\delta} & s_{23}c_{13} \\
   s_{12}s_{23}-c_{12}c_{23}s_{13}e^{i\delta} & -c_{12}s_{23}-s_{12}c_{23}s_{13}e^{i\delta} & c_{23}c_{13}
   \end{bmatrix}.
\label{eq:U}
\end{align}
Either a normal mass hierarchy (NH, $m_1 < m_2 < m_3$) or an inverted
mass hierarchy (IH, $m_3 < m_1 < m_2$) are allowed, following the
convention that the mass-squared difference $m_2^2 - m_1^2$ is responsible
for the oscillation of solar neutrinos. 

Suppose a model imposes a relationship among \textit{elements} of the
mass matrix $M$ given by
\begin{equation}
f\left( M_{\alpha \beta}\right)=0 \,,
\label{eq:elements}
\end{equation}
where $\alpha,\beta=e,\mu,\tau$ and $f$ is a homogeneous function of
the $M_{\alpha\beta}$. We take the coefficients in the
homogeneous function to be real, as in most models.
Then from Eq.~(\ref{eq:Mnu}),
$M_{\alpha\beta} = m_1V_{\alpha 1}^*V_{\beta 1}^* +
m_2V_{\alpha 2}^*V_{\beta 2}^* + m_3V_{\alpha 3}^*V_{\beta 3}^*$ and
Eq.~(\ref{eq:elements}) becomes
\begin{equation}
f\left(m_1V_{\alpha 1}^*V_{\beta 1}^*+m_2V_{\alpha 2}^*V_{\beta 2}^*+m_3V_{\alpha 3}^*V_{\beta 3}^*\right)=0 \,.
\label{eq:complex}
\end{equation}
Since the coefficients in the homogeneous function are real, the
complex conjugate of the above equation is
\begin{equation}
f\left(m_1V_{\alpha 1}V_{\beta 1}+m_2V_{\alpha 2}V_{\beta 2}+m_3V_{\alpha 3}V_{\beta 3}\right)=0 \text{~~~~~(element condition)}.
\label{eq:elements_condition}
\end{equation}

Now consider a model that imposes the same homogeneous relationship
among \textit{cofactors} of the light neutrino mass matrix, i.e.,
\begin{equation}
f\left( C_{\alpha \beta}\right)=0 \,,
\label{eq:cofactors}
\end{equation}
where $C_{\alpha \beta}$ is the $(\alpha, \beta)$ cofactor of $M$,
given by $(M^{-1})_{\alpha \beta}=\frac{1}{\det M}C_{\beta\alpha}$.
Since the mass matrix is symmetric and $f$ is a homogeneous function,
we have $f\left( (M^{-1})_{\alpha \beta}\right)=0$. Then since
$M^{-1}=V \text{diag}(m_1^{-1}, m_2^{-1}, m_3^{-1})V^T$, we can write
the condition as
\begin{equation}
f\left(m_1^{-1}V_{\alpha 1}V_{\beta 1}+m_2^{-1}V_{\alpha 2}V_{\beta 2}+m_3^{-1}V_{\alpha 3}V_{\beta 3}\right)=0\text{~~~~~(cofactor condition)}.
\label{eq:cofactors_condition}
\end{equation}

To compare the element NH case with the
cofactor IH case, we divide the argument in
Eq.~(\ref{eq:elements_condition}) by $m_3$, multiply the argument
in Eq.~(\ref{eq:cofactors_condition}) by $m_3$, and use the
properties of homogeneous functions to write the condition
for the element NH case as
\begin{equation}
f\left(\frac{m_1}{m_3} V_{\alpha 1} V_{\beta 1}+\frac{m_2}{m_3} V_{\alpha 2} V_{\beta 2}+V_{\alpha 3} V_{\beta 3}\right)=0 \,,
\end{equation}
and the condition for the cofactor IH case as 
\begin{equation}
f\left(\frac{m_3}{m_1} V_{\alpha 1} V_{\beta 1}+\frac{m_3}{m_2} V_{\alpha 2} V_{\beta 2}+V_{\alpha 3} V_{\beta 3}\right)=0 \,.
\end{equation}
In the quasi-degenerate regime ($m_1\simeq m_2\simeq m_3$), all the
mass ratios are approximately unity so that the three mixing angles
and three phases allowed by the constraints are nearly identical for
the two hierarchies.

For masses lighter than about 100 meV, the leading term in each
argument is the third term, and they are identical. The only
differences in the sub-leading terms are the two mass ratios. In the
figure we plot the fractional difference between the two mass ratios with
opposite hierarchies versus the lightest mass using the recent
best-fit values~\cite{Fogli:2012ua}, $\delta m^2 \equiv
m_2^2-m_1^2=7.54\times 10^{-5}\text{ eV}^2$ and $\Delta m^2 \equiv
|m_3^2-(m_1^2+m_2^2)/2|=2.43\times 10^{-3}\text{ eV}^2$ for the normal
hierarchy and $2.42\times 10^{-3}\text{ eV}^2$ for the inverted
hierarchy.  The percentage difference between $(\frac{m_1}{m_3})_{\rm
  NH}$ and $(\frac{m_3}{m_1})_{\rm IH}$ is very small and always less
than $1.7\% $ for any value of the lightest mass. The percentage
difference between $(\frac{m_2}{m_3})_{\rm NH}$ and
$(\frac{m_3}{m_2})_{\rm IH}$ becomes less than $10\%$ ($5\%$)
\{$2\%$\} if the lightest mass is larger than 19 (27) \{42\}
meV. Hence except for conditions with $\alpha$ and $\beta$ such that
$V_{\alpha 3} V_{\beta 3}$ is small compared to $V_{\alpha 1} V_{\beta
  1}$ and $V_{\alpha 2} V_{\beta 2}$, the two conditions are almost
the same for masses that are not nearly degenerate. Even in some extreme cases, such as $\alpha=\beta=e$, for which the $\theta_{13}$-dependent leading term is relatively small, the two conditions are almost identical if the lightest mass is larger than about 20 meV, with the percentage difference between the two mass ratios less than $10\%$.

\begin{figure}
\centering
\includegraphics[width=6.0in]{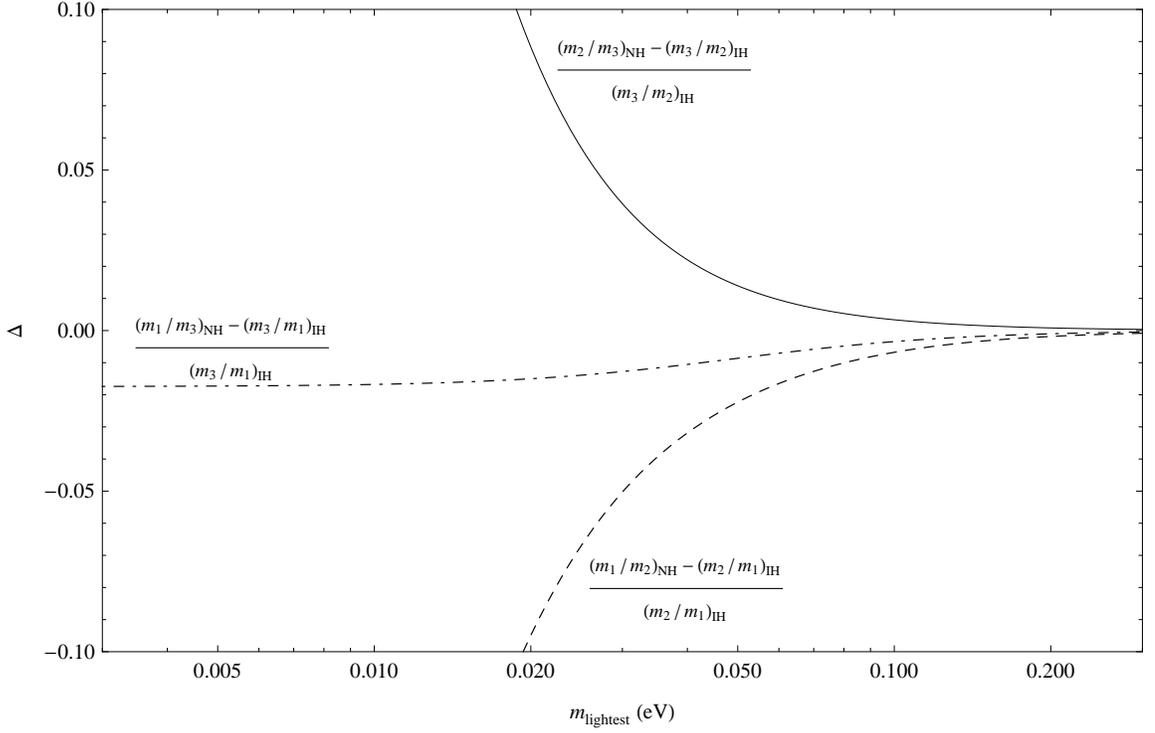}
\caption{Fractional differences in mass ratios $\Delta$ for the two mass
hierarchies as a function of the lightest neutrino mass. We set
$\delta m^2 \equiv m_2^2-m_1^2=7.54\times 10^{-5}\text{ eV}^2$ and
$\Delta m^2 \equiv m_3^2-(m_1^2+m_2^2)/2=2.43\times 10^{-3}\text{ eV}^2$
for the normal hierarchy and $2.42\times 10^{-3}\text{ eV}^2$ for the
inverted hierarchy.}
\label{fg:difference}
\end{figure}

To compare the element IH case with the cofactor NH case, we
divide the argument in Eq.~(\ref{eq:elements_condition}) by $m_1$
and multiply the argument in Eq.~(\ref{eq:cofactors_condition}) by
$m_1$ to obtain
\begin{equation}
f\left(V_{\alpha 1} V_{\beta 1}+\frac{m_2}{m_1} V_{\alpha 2} V_{\beta 2}+\frac{m_3}{m_1} V_{\alpha 3} V_{\beta 3}\right)=0 \,,
\end{equation}
for the element IH case and 
\begin{equation}
f\left( V_{\alpha 1} V_{\beta 1}+\frac{m_1}{m_2} V_{\alpha 2} V_{\beta 2}+\frac{m_1}{m_3}V_{\alpha 3} V_{\beta 3}\right)=0 \,,
\end{equation}
for the cofactor NH case. Again, in the quasi-degenerate regime
the two conditions are nearly identical, and the allowed values of the
mixing angles and phases are almost equal in the two models.

For masses lighter than about 100~meV, the leading terms in each
argument are the first two terms if the lightest mass is not very small. The percentage difference between $(\frac{m_2}{m_1})_{\rm
  IH}$ and $(\frac{m_1}{m_2})_{\rm NH}$ is less than $10\%$ ($5\%$)
\{$2\%$\} if the lightest mass is larger than 19 (30) \{53\}
meV. Hence, if the lightest mass is not very small, the two conditions
are almost identical. The sub-leading terms are always close to each
other because the percentage difference between $(\frac{m_1}{m_3})_{\rm
  NH}$ and $(\frac{m_3}{m_1})_{\rm IH}$ is always less than $1.7\%$
for any value of the lightest mass.

In the above analysis we only considered real coefficients in the
homogeneous functions. For complex coefficients, the complex conjugate
of Eq.~(\ref{eq:complex}) does not give
Eq.~(\ref{eq:elements_condition}). However, if the cofactor-based
model has coefficients that are the complex conjugate of the
coefficients in the element-based model, then the cofactor-based model
is dual to the corresponding element-based model.

Although our proof used only one condition, the same arguments can
easily be applied to multiple conditions.  The only requirement is
that there be two models with the same homogeneous conditions for the
elements and cofactors, respectively. A consequence is that a model
with two texture zeros yields predictions for the oscillation
parameters and Majorana phases similar to those of the corresponding
model with two cofactor zeros. Likewise, as noted in
Ref.~\cite{Dev:2013xca}, models with two equalities between mass
matrix elements are similar to models with two equalities between
cofactors.

{\bf Application to neutrino model building.}
The homogeneous relationships in Eqs.~(\ref{eq:elements}) and
(\ref{eq:cofactors}) are quite common in neutrino mass models, such as
texture zero models~\cite{Lashin:2011dn, Frampton:2002yf, texture
  zero}, cofactor zero models~\cite{Lavoura:2004tu, cofactor zero},
scaling models~\cite{Mohapatra:2006xy}, and models in which two mass
matrix elements or cofactors are equal~\cite{Dev:2013xca}. The latter
includes the $\mu-\tau$ symmetric models that impose $|M_{e\mu}| =
|M_{e\tau}|$ and $|M_{\mu\mu}|=|M_{\tau\tau}|$. However, the existence
of an element/cofactor duality requires models that have the same
homogeneous relationship among elements in one model and cofactors in
a second model. While models with conditions on the elements are
common, models with conditions on cofactors are not so common.
However, models with the same homogeneous relationships among
cofactors can be defined. In particular, the existence of the inverse
of the right-handed neutrino mass matrix in the conventional seesaw
mechanism~\cite{seesaw}, with $M=M_D^T M_R^{-1}M_D$, provides a good
motivation for the corresponding cofactor models, as we discuss below.

$M_D$ {\it is proportional to the unit matrix.}
A simple example arises when $M_D = m_D I$, so that inverting the
seesaw formula gives $M^{-1} = M_R/m_D^2$. Since $M^{-1} = C^T/{\rm
  Det}(M)$, it follows that $M_R \propto C^T$. Now since $M_R$ is
symmetric, any homogeneous relationship among the elements of the
right-handed neutrino mass matrix $M_R$ will be equivalent to the same
homogeneous relationship among the cofactors of the light neutrino
mass matrix. Thus a dual cofactor model can be obtained by having the
same homogeneous conditions on $M_R$ in one model as there are on $M$
in the dual element model; the cofactor conditions on $M$ are
inherited from $M_R$.

This leads to an even more ambitious conclusion: any model consistent
with the observed mixing angles (and phases) for the light neutrinos
will have a dual model with the opposite mass hierarchy. We can build the dual model by choosing $M_R$ to be proportional to $M$, and according to our argument above (with $M_D$ proportional to the unit matrix), the model generated by $M_R$ would
have a light neutrino cofactor matrix that is proportional to $M$. Thus
the cofactors are related to each other in the same way the elements
of $M$ are related to each other, so the model generated by $M_R$ with
the opposite mass hierarchy will be dual to the model represented by
$M$, and we cannot distinguish these two models without knowing the
mass hierarchy.

$M_D$ {\it is diagonal.}
Next we relax the condition that the Dirac mass matrix is proportional
to the unit matrix and consider the case where it is diagonal.  If we
assume that the same homogeneous relationship holds for both the light
neutrino mass matrix and the right-handed neutrino mass matrix, under
what conditions will the cofactor matrix of $M$ have the same
homogeneous relationship as the elements of $M$?

Defining $M_D=\text{diag} (c_1,c_2,c_3)$ and $(M_R)_{ij}=R_{ij}$,
since $M_R$ is symmetric, the cofactor matrix for $M$ becomes
\begin{align}
C = ({\rm Det} M) M_D^{-1} M_R^T (M_D^T)^{-1}
= ({\rm Det} M) \begin{bmatrix}
   R_{11}/c_1^2 & R_{12}/c_1c_2 & R_{13}/c_1c_3 \\
   R_{12}/c_1c_2 & R_{22}/c_2^2 & R_{23}/c_2c_3  \\
   R_{13}/c_1c_3 & R_{23}/c_2c_3 & R_{33}/c_3^2
   \end{bmatrix}.
\end{align}
We see that a texture zero in $M_R$ still translates to a cofactor
zero for $M$~\cite{Liao:2013rca, Ma:2005py}. However, in general a more
complicated homogeneous relationship among elements in $M_R$, such as
those involving more than one element, will not be inherited by the
corresponding cofactor matrix unless there is a special relationship
among the $c_i$. For example, $R_{\mu\mu} = R_{\tau\tau}$ does not
imply $C_{\mu\mu} = C_{\tau\tau}$ unless $c_2 = c_3$.

{\bf Conclusion.}
We have shown that if a model has a homogeneous relationship among
\textit{elements} of the light neutrino mass matrix, it will yield
predictions for the oscillation parameters and Majorana phases similar
to those of another model with the opposite mass hierarchy that has
the same homogeneous relationship among \textit{cofactors} of the mass
matrix, except when the lightest neutrino mass is of order
20~meV or less. Many existing models have one or more homogeneous
relationships among mass matrix elements, but there are fewer models
that are constructed by imposing homogeneous relationships among
cofactors. However, any model that fits current neutrino data will
have a dual model with the opposite mass hierarchy. We have shown that
if the Dirac mass matrix is proportional to the identity matrix, a
dual cofactor-based model can be generated via the seesaw mechanism if
the right-handed neutrino mass matrix has the same homogeneous
relationships as the light mass matrix elements in an element-based
model. Since the mass hierarchy has not been experimentally
determined, we cannot currently distinguish these dual models from each
other.

Current global fits to oscillation data have almost identical best-fit
values for the neutrino mixing angles and mass-squared differences for
the two hierarchies. However, different allowed regions for the
oscillation parameters for different mass hierarchies can lead to a
breakdown of duality. In fact, the $2\sigma$ allowed regions are
somewhat different, especially for the value of $\theta_{23}$, where
second octant values ($\theta_{23} > \pi/4$) are allowed only for the
inverted hierarchy~\cite{Fogli:2012ua}. Due to this difference, the
$2\sigma$ allowed regions for dual models differ even in the
quasi-degenerate regime in a few cases we have
studied. 

The dual model ambiguity can be resolved by experiments that distinguish
between the normal and inverted hierarchies, such as long baseline neutrino experiments (T2K~\cite{Abe:2011ks}, NO$\nu$A~\cite{Patterson:2012zs}, and LBNE~\cite{Akiri:2011dv}), atmospheric neutrino experiments (PINGU~\cite{Clark:2012hya}, and INO~\cite{Mondal:2012xm}) and medium baseline reactor experiments (Daya Bay II/JUNO~\cite{dayabayii:dayabayii, Kettell:2013eos}), or a combination of these~\cite{Blennow:2013vta}.  
Also, tritium beta decay, neutrinoless double beta decay ($0\nu \beta \beta$), and structure formation in our universe  depend on the nature of the neutrino mass pectrum, and in principle can be used to distinguish between dual models.  The 95\% C.L. sensitivity of the KATRIN experiment~\cite{Osipowicz:2001sq} to the effective neutrino mass $m_\beta=(\sum\limits_i |V_{ei}|^2 m_i^2)^{1/2}$ is 0.35~eV with an uncertainty of 0.08~eV$^2$ on $m_\beta^2$, which is insufficient to break the duality. 
 The effective Majorana mass measured by
 $0\nu \beta \beta$ experiments is constrained to be smaller than 140-380 meV at the 90\% C.L.~\cite{Auger:2012ar}, which cannot break the duality. The future sensitivity of $0\nu \beta \beta$ experiments is expected to be 50~meV or lower~\cite{Rodejohann:2012xd}, which would provide a partial but strongly model-dependent resolution of the dual model ambiguity~\cite{Liao:2013kix,Liao:2013rca}.
 The current 95\% ~C.L. upper bound on $\Sigma m_i$ from cosmology is 0.66~eV~\cite{Reid:2009xm}, which permits a quasi-degenerate spectrum, so that the duality is unbroken.  In the future, lensing measurements will probe $\Sigma m_i$ as low as 0.05 eV~\cite{dePutter:2009kn}, which will help distinguish between dual models. 


{\bf Acknowledgments.}
DM thanks the Kavli Institute for Theoretical Physics for its
hospitality while this work was in progress. KW thanks the Center for
Theoretical Underground Physics and Related Areas (CETUP* 2013) in
South Dakota for its hospitality and for partial support while this
work was in progress. This research was supported by the
U.S. Department of Energy under Grant Nos. DE-FG02-01ER41155,
DE-FG02-04ER41308, and DE-FG02-13ER42024, and by the National Science
Foundation under grant No. PHY11-25915.

\end{document}